\documentclass[aps,prl,reprint,groupedaddress]{revtex4-1}
\usepackage{graphicx}

\def\ta0{\tilde a_0}
\def\ta{\tilde a}
\def\be{\begin{equation}}
\def\ee{\end{equation}}
\def\bc{\begin{columns}}
\def\ec{\end{columns}}
\def\bea{\begin{eqnarray}}
\def\eea{\end{eqnarray}}

\begin{document}


\title{Differences in mechanical properties lead to anomalous phase separation in a model cell co-culture}



\author{Supravat Dey}
\email{supravat.dey@gmail.com}
\thanks{\emph{Present address:}Department of Electrical and Computer Engineering, University of Delaware, Newark,  DE 19716, USA.}
\affiliation{School of Physics and Astronomy, Rochester Institute of Technology, Rochester, New York 14623, USA.}
\author{Moumita Das }
\email{modsps@rit.edu}
\affiliation{School of Physics and Astronomy, Rochester Institute of Technology, Rochester, New York 14623, USA.}


\date{\today}

\begin{abstract}
During the morphogenesis of tissues and tumors, cells often interact with neighbors with different mechanical properties, but the understanding of its role is lacking.  We use active Brownian Dynamics simulations to study a model co-culture consisting of two types of cells with the same size and self-propulsion speed, but different mechanical stiffness and cell-cell adhesion. As time evolves, the system phase separates out into clusters with distinct morphologies and transport properties for the two cell types. The density structure functions and the growth of cell clusters deviate from behavior characteristic of the phase separation in binary fluids. Our results capture emergent structure and motility previously observed in co-culture experiments and provide mechanistic insights into intercellular phase separation during development and disease.
\end{abstract}


\maketitle


Self-organization and phase separation of cells are critical to tissue morphogenesis, whether in the context of formation of embryos or the formation and progression of tumors. During these processes, cells are often surrounded by neighbors with different mechanical properties. For example, recent studies have shown that for many types of tumors, cancer cells are mechanically softer and more deformable than the corresponding non-cancerous cells \cite{Suresh,Lekka2012,Lee2012}.  Furthermore, while healthy, epithelial cells express specific proteins such as E-cadherin that facilitate cell-to-cell adhesion, in most cancer cells and other mesenchymal cells, the expression of this protein is significantly down regulated \cite{Oka,Wijnhoven,Ma2016,Wong}. This leads to the following questions: how do the variations in the mechanical  properties of cells influence their collective dynamics and self-organization?  Can some cells harness these differences for enhanced organization and migration better than others? Addressing these questions will help to understand the role of mechanical heterogeneity in cellular organization, dynamics, and functions, and may provide insights into how changes in cells' mechanical properties during tumorigenesis influence cancer invasion and metastasis. 

In recent years, there has been tremendous focus on phase separation within cells, the associated physical mechanisms, and the consequences for biological function  \cite{Hyman,Brangwynne,Shin,Alberti}. When it comes to phase separation of multicellular systems, while the fully segregated final states can be understood in terms of the Differential Adhesion Hypothesis (DAH) \cite{Steinberg1962,Steinberg1994,Steinberg2005,Manning2010,Brochard,Ma2016} as corresponding to the minimum in the free energy of a mixture of immiscible liquids with different surface energies, the dynamics via which these states are reached is not well understood. Furthermore, the DAH focuses on the differences in cell adhesion and does not account for the differences in cell stiffness. During the past decade, there have also been many computational studies of phase separation in single species active colloidal systems \cite{Buttinoni2013,Palacci2013,Fily,Redner,Cates2015} and in binary systems motivated by mixtures of cells with different motility or adhesion  \cite{Belmonte2008,Mehes2012,Beatrici2017,Graner1992,Kabla2012,Nakajima2011,Osborne2017}. In the latter, the interactions between cells were phenomenological, and an understanding of the role of differences in cell mechanical properties is still lacking. Here we ask how do differences in cell stiffness and adhesion impact their phase separation in a co-culture and whether the structure and dynamics of the phase separated system have characteristics similar to binary liquids. 

\begin{figure}[th]
\includegraphics[width=0.5\textwidth]{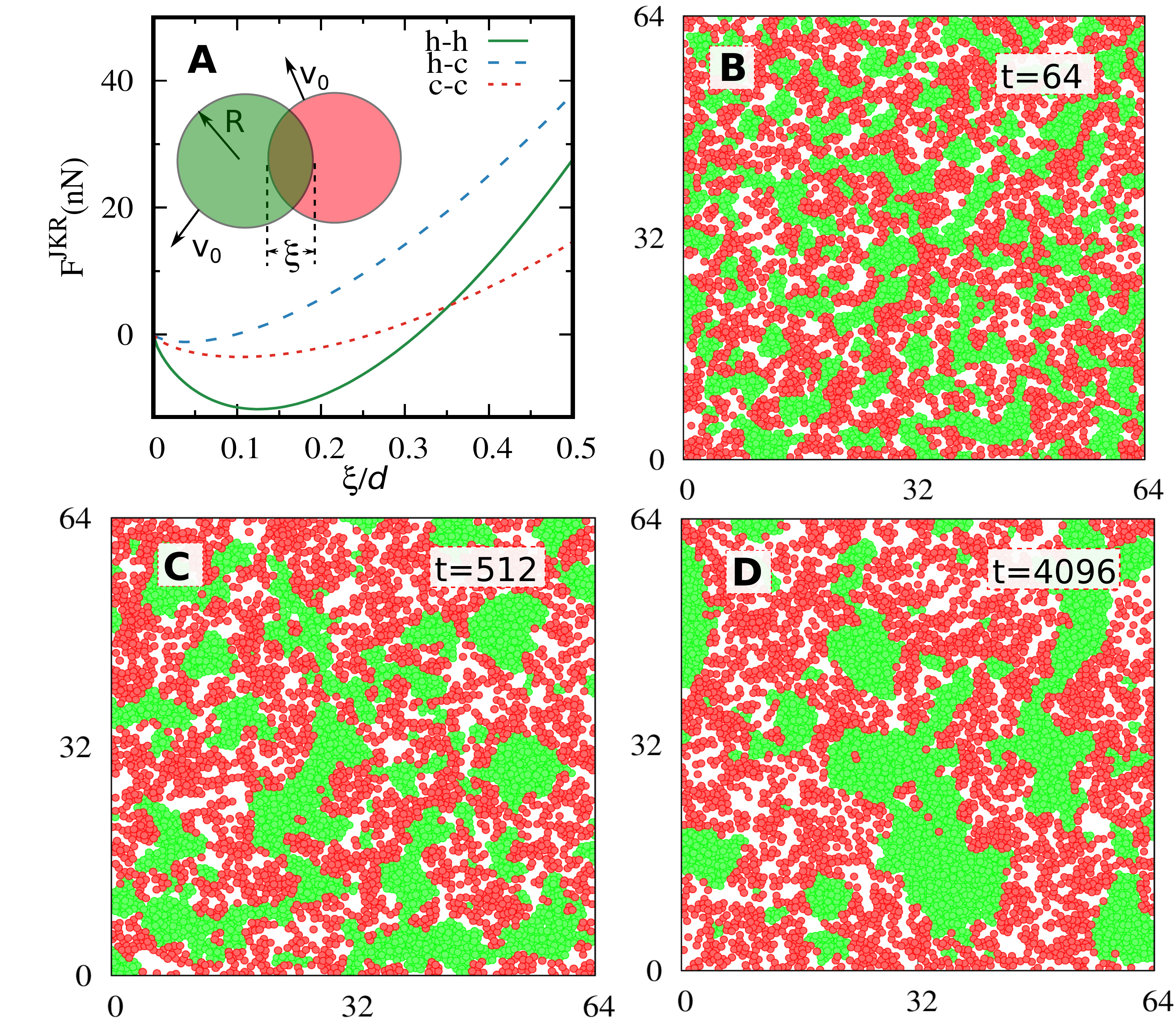}
\caption{A. The JKR forces between two cells for $h-h$, $h-c$, and $c-c$ contacts  as a function of scaled overlap distance for parameters in Table 1. The forces are attractive at small and 
repulsive at large overlap distances. The attractive force is strong for $h-h$ contact and weak for $c-c$ contacts. B-D show phase separation of $h$- (green) and $c$- (red) cells at time $t=64, 512$, and $4096$, starting from an initially homogeneous mixture. Images show subsets (size $64\times64$) of the system.}
\label{fig:snap}
\end{figure}

To address these questions, we study a model co-culture consisting of two different types of cells that have the same size and self-propulsion speed but different Young's moduli and surface energies. We model this system as a 2D suspension of spherical overdamped self-propelled particles with mechano-adhesive contact interactions. The equation of motion of a cell $i$ $\in\{1,N\}$ ($N$ being the total number of cell) with position vector $\vec{r}_i$ and the orientation angle $\theta_i$ of self-propulsion is given by \cite{Fily,Redner},
\bea
\frac{d\vec{r}_i}{dt}=v_0 \hat{n}_{\theta,i} + \mu \sum_{j\in n_i}\vec{F}(\vec{r_{i}}, \vec{r_{j}}), \;\;\;\;
\frac{d\theta_i}{dt}=\sqrt{2\,D_r}\eta_{_i},
\label{eqn:pos}
\eea
where $\hat{n}_{\theta,i}$ and $v_0$ are the unit vector and the speed associated with the propulsion velocity, and $\mu$ is the mobility. 
The intercellular force between cells $i$ and $j$, $\vec{F}(\vec{r_{i}}, \vec{r_{j}})$, is non-zero only when the two cells are in contact, and the sum is over all contacting neighbors of cell $i$. 
Here, we neglect the translation noise as in \cite{Szabo2006,Belmonte2008,Smeets2016}. The random variable $\eta_i(t)$ has zero mean and is $\delta-$ correlated i.e., $\langle \eta_i(t) \rangle = 0$ and $\langle \eta_i(t)\eta_j(t^\prime) \rangle = 2\,\delta_{i,j}\delta(t-t^\prime)$ and $D_r$ is the angular diffusivity.  Furthermore, there is no built-in alignment interaction between cells; these interactions may impact the collective dynamics of some cell types \cite{Mehes2013}, and will be incorporated in future work.


We model the contact force between cells using the well known Johnson Kendall Roberts (JKR) model \cite{Johnson1971}. The JKR model was originally proposed for soft sticky spherical objects in contact \cite{Johnson1971}, and has been experimentally shown to work well for several cell types \cite{Chu2005, Zhu2016}. Given constitutive cell mechanical properties such as cell stiffness and interfacial energy of cell-cell contacts, the JKR model gives the intercellular contact force as a function of the overlap distance between cells, and is therefore better suited for studying differences in mechanical properties than commonly used phenomenological models which assume the contact force to be a piecewise linear function of the intercellular distance \cite{Szabo2006,Belmonte2008,Smeets2016}. 


For a pair of overlapping spherical particles $i$ and $j$, the JKR force can be formulated in terms of the virtual overlap $\xi_{ij}=d - |\vec{r}_{ij}|$, where 
$\vec{r}_{ij}=\vec{r}_j - \vec{r}_i$ and $d$ is the particle diameter (Fig.~\ref{fig:snap}A). For computational simplicity, we use an approximate form of this force which assumes weak adhesion and is given by \cite{Brilliantov2005,Beyer2009},
\bea
\!\!\vec{F}^{\rm JKR}_{ij} = \frac{1}{2} \bigg[ Y^{\rm eff}_{ij}{d}^{1/2}\xi_{ij}^{3/2} 
\!-\!\sqrt{3 \pi \sigma_{ij} Y^{\rm eff}_{ij}} {d}^{3/4}\xi_{ij}^{3/4} \bigg] \frac{\vec{r}_{ij}}{|\vec{r}_{ij}|},
\label{eqn:fjkr}
\eea
where $\sigma_{ij}$ is the interfacial energy of the contact between $i$ and $j$, and $Y^{\rm eff}_{ij}$ is their effective Young modulus, given by
$\frac{1}{Y^{\rm eff }_{ij}}=  \frac{3}{4}\left( \frac{1-\nu^2_i}{Y_i} +  \frac{1-\nu^2_j}{Y_j} \right)$.
Here, $Y_{i(j)}$ and $\nu_{i(j)}$ are the Young's modulus and Poisson's ratio of particle $i(j)$.

The first term in Eq.~\ref{eqn:fjkr} corresponds to a repulsive contribution due to cell stiffness, and the second term to cell-cell adhesion. The equilibrium overlap distance between the two particles is given by the value of $\xi_{ij}$ where $F^{\rm JKR}=0$, and the force required to pull the two particles off each other is the maximum negative value of the JKR force (See Fig.~\ref{fig:snap}A). Both these quantities are critical to the collective structure and dynamics of the system and depend on the elasticity of the particles and the interfacial energy of the contact. While the JKR model has been used in several other theoretical and computational studies of cell mechanics \cite{Drasdo2007,Byrne2008,Brilliantov2005,Beyer2009}, to our knowledge it has not been used to study demixing in cell co-cultures. 

With these model ingredients, we consider two cell types, which we refer to as $h$ and $c$ cells, with distinct mechanical properties. The $h$ and $c$ cells are motivated by the healthy breast epithelial cell line MCF10A and the breast cancer cell line MDA-MB231 respectively, and have approximately the same Young's moduli and ratio of surface energies as these cells (see Table \ref{tab:table1}). Therefore,  $Y_h > Y_c$, and  $\sigma_{hh}>\sigma_{cc}>\sigma_{hc}$. The resulting JKR force exhibits stronger adhesion at small overlap distances and consequently a larger pull-off force, as well as larger elastic repulsion at large overlap distances for healthy breast cells ($h-h$) than breast cancer cells ($c-c$) (Fig.~\ref{fig:snap}A). 

To simulate the dynamics of this model system, we first non-dimensionalize the dynamical equations ( Eqs.~\ref{eqn:pos}) using cell diameter $d$ as the characteristic length scale, and time $\tau=d/v_0$ needed to cross $d$ when moving with a self-propulsion speed $v_0$ as the time scale. We integrate these equations using the Euler method with a step size $\delta t=0.001\,\tau$. The simulations are performed in a 2D box of size $L\,d\times L\,d$ with equal numbers of h and c cells for area fraction $\phi$, using periodic boundary conditions. 

\begin{table}[h!]
  \begin{center}
    \begin{ruledtabular}
    \begin{tabular}{|c|c|c|c|c|c|} 
	    $v_0$ & 1 $\mu$m min$^{-1}$\cite{Smeets2016} & $D_r$ & 0.05\,min$^{-1}$\cite{Smeets2016} & $\mu^{-1}$ & 8 nN$\mu$m min$^{-1}$ \\
      \hline
	    $Y_{h}$ & 1.5 nN$\mu$m$^{-2}$\cite{Lekka2012} & $Y_{c}$ & 0.4\,$Y_{h}$\cite{Lee2012} & $\nu$ & 0.5  \\
      \hline
	    $\sigma_{hh}$ & 0.5 nN$\mu$m$^{-1}$\cite{Heisenberg2013} & $\sigma_{cc}$ & 0.3\,$\sigma_{hh}$ & $\sigma_{hc}$ & 0.1 $\sigma_{hh}$ \\
      \hline 
	    $d$ & 20 $\mu$m & $L$ & 256 & $\phi$ & 1.2 \\
    \end{tabular}
    \end{ruledtabular}
        \caption{Simulation parameters}
    \label{tab:table1}
  \end{center}
\end{table}

\begin{figure}[h!]
\includegraphics[width=0.5\textwidth]{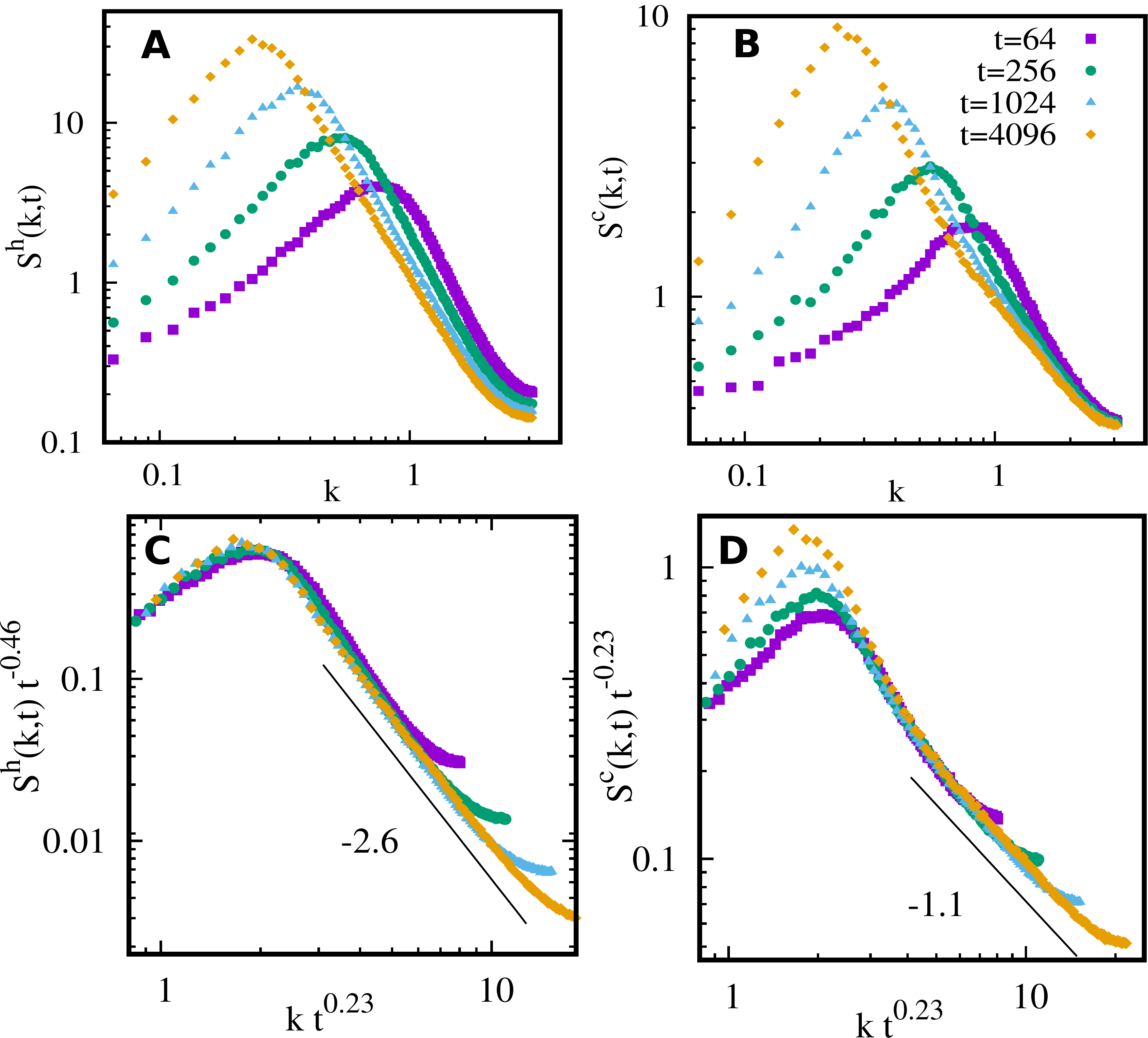}
\caption{A and B show the density structure factors for $h$ and $c$ cells, respectively, as a function of the wavevector $k$ at times $t=64, 256, 1024$, and $4096$.  C and D show the scaling collapse of the above data according to Eq.~\ref{eqn:scaling_hypothesis} with $\mathcal{L}(t)\sim t^{0.23}$. C.  The scaled data for $h$ cells show a power law decay with exponent $-2.6$ for large $k\mathcal{L}(t)$. D.  For $c$ cells, the data collapse is possible only at large $k$, and with d replaced by $d/2$; the decay exponent is now $-1.1$.} 
\label{fig:sfactor}
\end{figure}


Starting with a spatially homogeneous mixture, we find that as time evolves, the $h$ and $c$ cells separate out, forming clusters of individual cell types which coalesce and grow. The time evolution of a typical phase-separated pattern is shown in Fig.~\ref{fig:snap}B-D. The clusters of the two cell types have distinct morphologies: while $h$ cell clusters are compact with well-defined boundaries, those of $c$ cells are dispersed with ill-defined boundaries. Similar morphologies were observed in recent experimental studies of co-cultures of MCF10A and MDA MB231 \cite{Wong,Pawlizak15}. 

Here we note that unlike \cite{Beatrici2011} where phase separation is driven by a difference in self-propelled velocity, in our model all the cells have the same self-propulsion speed and there is no alignment interaction between them. Coming to comparisons with binary liquids, interfacial tension plays a critical role in the phase separation of our model co-culture as in these systems. However, the compact and dispersed morphologies seen here are not usual in binary liquids \cite{Bray1994}. As far as we are aware, they have not been reported in theoretical studies of phase separation before.  The origin of such distinct cluster structures lies in the interplay of differences in mechanical properties of the cells. The combination of greater stiffness and greater cell-cell adhesion of $h$ cells  (Fig.~\ref{fig:snap}A) causes them to form clusters that are simultaneously more closely packed and have a well-defined boundary. On the other hand, $c$ cells are less adhesive and less stiff, leading to both weak attraction when they come in contact and weak repulsion when they push each other, resulting in dispersed clusters with irregular boundaries. Together with the barely attractive $h-c$ interactions, this leads to rough interfaces between the $h$ and $c$ cell domains. 


To characterizes the morphologies of the clusters and growth, we measure spatio-temporal density structure factors. First, we define a local coarse-grained density field for a cell-type $\alpha$, $\rho^{\alpha}(\vec{r},t)$ as the total number of $\alpha$ ($=h$ or $c$) cells inside a square box of unit area. Then, the density structure factor is given by $S^{\alpha}_{\rho \rho}(\vec{k}, t) = \langle \tilde\rho^{\alpha}(\vec{k},t)\tilde \rho^{\alpha}(-\vec{k},t) \rangle$,
where $\tilde\rho^{\alpha}(\vec{k},t)$ is the Fourier transform of $\rho^{\alpha}(\vec{r},t)$ at a wave vector $\vec{k}$  and the angular bracket represents average over many ensembles.   

For a system undergoing phase ordering kinetics, the structure factor follows the scaling hypothesis \cite{Bray1994}:
\begin{eqnarray}
S{\rho \rho}(k, t)&=& \mathcal{L}^d(t) g(k\mathcal{L}(t)),
\label{eqn:scaling_hypothesis}
\end{eqnarray}
where $\mathcal{L}(t)$ is the growing length scale for the density field and $d$ is the dimensionality. The scaling function $g(y)$ shows a power law decay as $g(y) \sim y^{-\theta}$ for $y\gg 1$. For a system with a sharp, regular interface, the structure factor of a scaler order parameter, such as density, has an exponent $\theta=d+1$. This is the well known Porod law \cite{Porod1984}. Phase separation in binary fluids ordering towards equilibrium often show morphologies obeying this law \cite{Bray1994}. Deviations from the Porod law may indicate a rough interface or structural hierarchy and have been observed in non-equilibrium systems \cite{Das00,Shinde07,Mishra06,Dey12}. The length scale of typical  $\mathcal{L}(t)$ grows as $t^\beta$, where $\beta=1/3$ for a system with a conserved order parameter and is attributed to growth dominated by diffusion such as in diffusive coalescence of droplets in binary liquids \cite{Bray1994,Siggia,Furukawa}. 

We measure the structure factor for $h$ and $c$ cell clusters, separately. In Fig.~\ref{fig:sfactor}A and B, we show the  structure factor data $S^{(h)}_{\rho\rho}$ for $h$ cells  and
$S^{(c)}_{\rho\rho}$ for $c$ cells respectively at four different times. We found that for $h$ cells, the data at different times show an excellent collapse according the scaling hypothesis (Eq.~\ref{eqn:scaling_hypothesis}) with $\mathcal{L}(t)\sim t^{\beta_h}$, where $\beta_h\simeq0.23\pm0.02$ (Fig.~\ref{fig:sfactor}C). The large $k\mathcal{L}(t)$ data of scaled structure factor, show a power law decay with exponent $\theta^{h}=2.6$. This is a small but clear deviation from the Porod law which predicts an exponent $3$ in $2D$. We believe it arises because the interfaces between the $h$ and $c$ domains are rough even though the $h$ domains themselves are compact. The structure factor for $c$ cell clusters are distinct from $h$ cell clusters. For $c$ cells, we are able to obtain a good scaling collapse of the structure factor data only at large wavevectors, and only when we replace $d$ in equation (Eq.~\ref{eqn:scaling_hypothesis}) by $d/2$, or in other words when the structure factor is additionally scaled with $t^{-0.23}$, an unusual scaling which was previously observed in a freely cooling granular gas \cite{Shinde07}.  At large wavevectors, this scaled structure factor shows a power law decay with an exponent $\theta^{c}=1.1$, indicating considerably stronger violation of the Porod law due to the dispersed nature of the $c$ cell clusters in addition to the rough $h-c$ phase boundaries. 

Also, note the deviation of the cluster growth exponent $\beta$ from the scaling $1/3$ commonly observed in droplet coalescence in liquid-liquid phase separation. This may be qualitatively understood as follows. In binary liquids, droplet merging due to Brownian diffusion gives a timescale of growth $t \sim {\mathcal{L}(t)}^2/D$, where the diffusivity $D \sim K_B T/\eta \mathcal{L}(t) $, which in turn gives $ \mathcal{L}(t) \propto t^{1/3}$ \cite{Bray1990,Siggia}. In the above, $K_B$ is the Boltzmann constant, $T$ the temperature, and $\eta$ the viscosity. In our model co-culture, the effective diffusivity underlying the coarsening of clusters scales as $1/ \mathcal{L}(t)^d$ in the absence of alignment interactions between cells \cite{Beatrici2017,Kolb1984}, which leads to a growth law $ \mathcal{L}(t)  \propto t^\frac{1}{d+2}$ ($t^{1/4}$ in 2D) \cite{Furukawa}. A growth exponent  $\sim 1/4$ was previously reported in some passive \cite{Furukawa,Coniglio1995,Wolterink2006} and active phase separating systems \cite{Dey12,Redner2013,Beatrici2017,Stenhammar2014}.   
 
\begin{figure}[!t]
\includegraphics[width=0.5\textwidth]{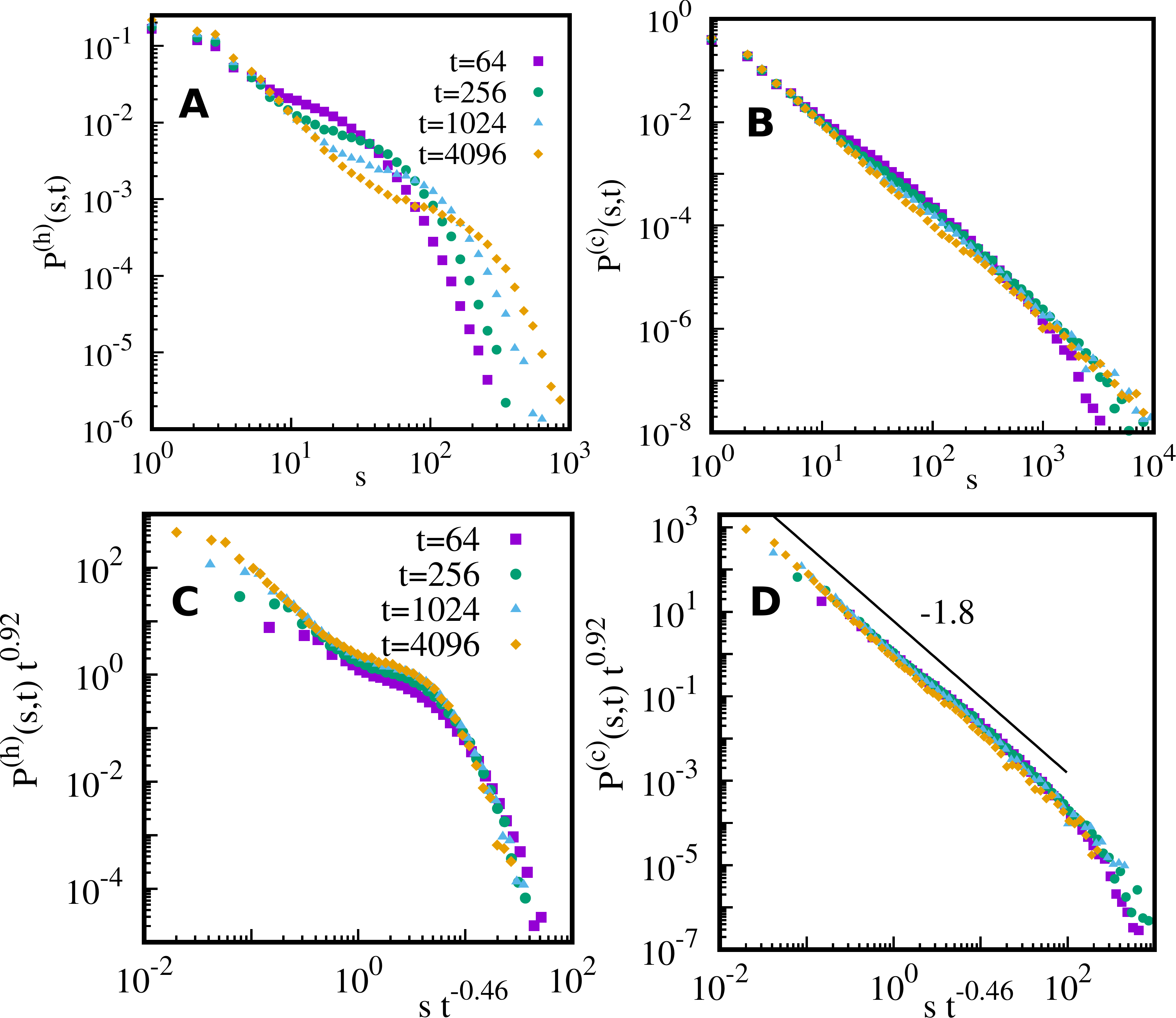}
\caption{A and B show the probability distributions of clusters of $h$ and $c$ cells, respectively, as a function of cluster size $s$, at times $t=64, 256, 1024$, and $4096$. 
C and D show the collapse of cluster size distribution data on to single curves  for $h$ and $c$ cells according to Eq.~\ref{fig:sdist} respectively. The latter further shows a power law decay with exponent $-1.8$.}  
\label{fig:sdist}
\end{figure}

To understand the above morphologies and growth laws better, we measure the distribution of cluster sizes $P^{(\alpha)}(s,t)$ at different times, where $s$ is the total number cells in a cluster of cell type $\alpha$. In Fig.~\ref{fig:sdist}, we plot the probability distribution for cluster sizes for different times for $h$ and $c$ cells. We observe a data collapse according to,
\bea
P^{\alpha}(s,t) = \frac{1}{\mathcal{A}(t)^2} f^{\alpha} (s/\mathcal{A}(t)),
\eea   
where $\mathcal{A}(t)=\mathcal{L}(t)^{d}=t^{2\beta}=t^{0.46}$ which implies that cluster sizes for both $h$ and $c$ cells grow as $t^{0.23}$, consistent with the structure factor measurement. For $c$ cells, we observe that the cluster sizes are distributed as a power law over many decades (Fig.~\ref{fig:sdist}D), suggesting hierarchical clusters as the origin of the non-Porod behavior. We find the decay exponent for cluster size distribution is $-1.8$. The implication of this particular exponent will be explored in future work. For $h$ cells, we do not observe a clear power law decay of clusters sizes.  
  
\begin{figure}[!t]
\includegraphics[width=0.5\textwidth]{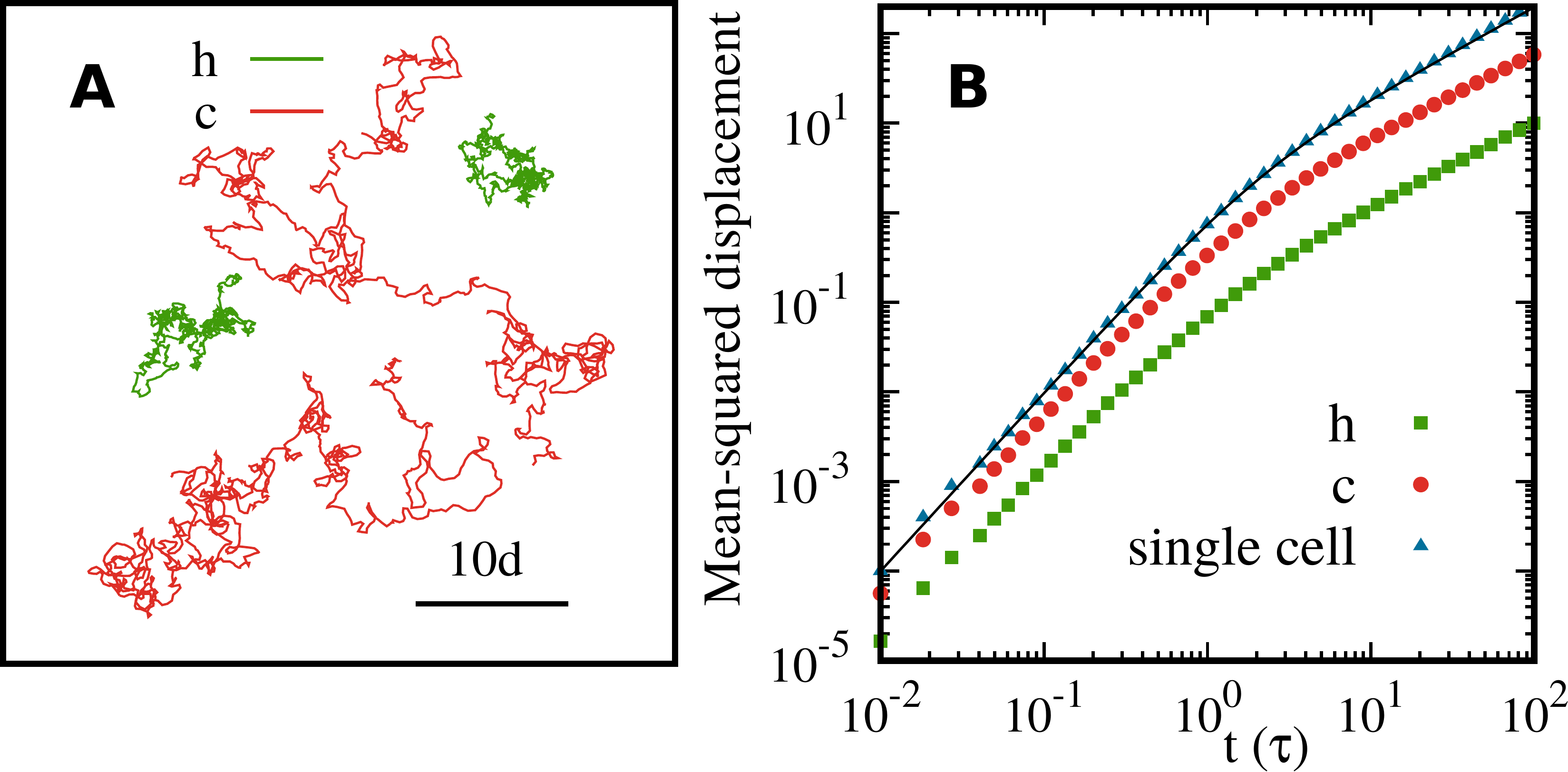}
\caption{A. Sample trajectories of $h$ and $c$ cells in a co-culture show that $c$ cells are more motile than $h$ cells. B shows that $c$ cells have a larger MSD than $h$ cells in a co-culture, but both have a lower MSD than a single cell in isolation.} 
\label{fig:msd}
\end{figure}

Finally, we turn our attention to the emergent motility of $h$ and $c$ cells. We measure the mean squared displacement (MSD) for the $h$ and $c$ cells once they reach  a ``steady state'' at time $t_0$. As the phase separation dynamics is very slow, for computational efficiency, we calculate MSDs for a relatively small system size $L=64\,d$, and  $t_0=8192$ (Fig.~\ref{fig:msd}). The MSD for  a single isolated cell with self-propelled speed $v_0$ and angular noise strength $D_r$ is given by, $\langle\delta r^2(t)\rangle = 4 D_0\left[t +  1/D_r({\rm e}^{ D_rt}-1\right)]$, where $D_0=v^2_0/(2 D_r)$ \cite{Szabo10}. The motion is ballistic at small  times scales and diffusive at large timescales. We observe a similar ballistic to diffusive crossover in the model co-culture, but with a smaller effective diffusion coefficient and effective velocity than for a single isolated cell. More interestingly, the $h$ and $c$ cells follow two different MSD curves although their self-propulsion speeds are the same, with the MSD of the $c$ cells being greater than that of the $h$ cells at any given time. This suggests that the effective motility of $c$ cells in a co-culture is greater than that of the $h$ cells, consistent with experimental observation \cite{Lee2012}, and can arise solely from the differences in mechanical properties of the two cell types.

In summary, we have studied the phase separation of a model co-culture of two cell types $h$ and $c$, where $c$ cells are mechanically softer and show weaker cell-cell adhesion than $h$ cells. We used previously reported experimental values of breast epithelial cells and breast cancer cells for the mechanical properties of the $h$ and $c$ cells respectively. In numerical simulations of this system where we modeled the cells as active particles interacting via mechanical contact forces, we observed distinct morphologies of the phase separated domains qualitatively similar to those observed in experiments \cite{Wong,Pawlizak15}; while $h$ cells formed compact clusters, $c$ cell domains were much more dispersed. We found that the structure functions of both $h$ and $c$ cell clusters deviate from the Porod behavior. The deviation was particularly striking for the $c$ cell clusters and arise due to a hierarchy of cluster sizes. We note that the non-Porod behavior has been observed in other biological contexts such as the microstructure of goat femur \cite{Rajesh}. The $c$ cells also exhibited enhanced emergent motility aided by bursty movement consistent with previous experimental observations \cite{Lee2012,Wong}. The domain coarsening follows a $t^{1/4}$ growth law due to slower coalescence of clusters. Our results demonstrate how mechanical heterogeneity can lead to phase separation and differential cell motility in co-cultures of epithelial and mesenchymal cells, and may provide insights into the roughness of tumor spheroids and enhanced migration of cancer cells during tumor progression.

\begin{acknowledgments}
The authors thank D. Das for helpful discussions and comments on the manuscript, and J. Feng and C. Beatrici for informative conversations. 
 This research was funded in part by the Gordon and Betty Moore Foundation through Grant GBMF5263.02 and a Cottrell College Science Award from the Research Corporation.
\end{acknowledgments}


%

\end{document}